\documentclass{aa}
\usepackage{graphicx}
\usepackage{latexsym}
\usepackage{epsfig}
\usepackage{longtable}
\usepackage{txfonts}
\begin{document}
   \title{The correlation between soft and hard X-rays component in
  flares: from the Sun to the stars}

   \author{C. Isola
          \inst{1}
          \and
          F. Favata\inst{1}
          \and
          G. Micela \inst{2}
      \and
          H.  S. Hudson \inst{3}
          }

   \institute{ESA-ESTEC, Astrophysics Division - Research and Scientific Support Department,
              Postbus 299, 2200 AG, Noordwijk ZH, The Netherlands\\
              \email{cisola@rssd.esa.int; ffavata@rssd.esa.int}
         \and
             INAF, Osservatorio astronomico di Palermo,
             Piazza del parlamento 1, 90134 Palermo, Italy \\
             \email{giusi@astropa.inaf.it}
     \and
             SSL/UCB, 7 Gauss Way, Berkeley,
           CA 94720-7450, United States \\
         \email{hhudson@ssl.berkeley.edu}
             }
% 5 {} token are mandatory

  \abstract
    {}
   {In this work we study the correlation between the soft (1.6--12.4 keV, mostly thermal) and the hard (20--40 and 60--80 keV, mostly non-thermal) X-ray emission in solar flares up to the most energetic events, spanning about 4 orders of magnitude in peak flux, establishing a general scaling law and extending it to the most intense stellar flaring events observed to date.}
 {We used the data from the Reuven Ramaty High-Energy Solar Spectroscopic Imager (RHESSI) spacecraft, a NASA Small Explorer launched in February 2002.
RHESSI has good spectral resolution ($\simeq 1$~keV in the X-ray range) and broad energy coverage (3~keV--20~MeV), which makes it well suited to distinguish
the thermal from non-thermal emission in solar flares. Our study is based on the detailed analysis of 45 flares ranging from the GOES C-class, to the strongest X-class events, using the peak photon fluxes in the GOES  1.6--12.4 keV and in two bands selected from RHESSI data, i.e.\
20--40 keV and 60--80 keV.}
{We find a significant correlation between the soft and hard peak X-ray fluxes spanning the complete sample studied.
The resulting scaling law has been extrapolated to the case of the most intense stellar flares observed, comparing it with the stellar observations.}
 % conclusions heading (optional)
{{ Our results show that an extrapolation of the scaling law derived for 
solar flares to the most active stellar events is compatible with the
available observations of intense stellar flares in hard X-rays.}}


\maketitle
%
%________________________________________________________________

%\section{Introduction}
%__________________________________________________________________

\section{Introduction}

Solar flares are violent explosions taking place
in the solar corona and chromosphere, generally attributed to magnetic reconnection events.
They result in the rapid release of a large amount of energy, up to $10^{32}$ ergs, in $10^{2}$--$10^{3}$ s.

A variety of phenomena is associated with the sudden energy release in a flare, including the acceleration of particles (electrons up to 
tens of MeV and ions up to tens of GeV) and the heating of coronal plasma to  10--30 MK.
The radiation emitted by a solar flare covers
the entire electromagnetic spectrum,
from radio waves to X-rays and $\gamma$-rays, but
the different wavelengths are associated with
different regions of the solar atmosphere and
with different mechanisms, making it necessary to take a ``multi-wavelength'' 
approach to the study of flare physics.

In spite of the large number of solar flare observations  performed with many instruments, the detailed physical model associated with a flare remains elusive.
The ``non-thermal thick-target model'' (Brown
1971; Hudson 1972; Lin $\&$ Hudson 1976), 
in which a beam of electrons accelerated at the site of magnetic reconnection (in the corona), are braked by the the cooler chromospheric plasma (the ``target''), producing the observed hard X-rays emission, has most commonly been used to explain many of the observed properties in solar flares.

In this framework the flare has an {\it impulsive phase} and a {\it gradual phase}. In the impulsive phase the electrons
are accelerated to weakly relativistic energies and impact on the chromosphere, causing the heating and the evaporation of plasma into a coronal loop. The impulsive phase (lasting at most a few minutes in the Sun) is therefore characterized by hard X-rays of non-thermal origins. The gradual phase (which in the Sun can last from a few minutes to several hours) is characterized by the soft X-ray emission which is (together with conduction to the chromosphere) the main cooling mechanism of the evaporated plasma filling a coronal loop.

In this model only a small fraction of the energy of the accelerated electrons is lost
through radiation: most of the loss is due to Coulomb collisions heating the chromospheric plasma, provoking its rapid heating at the loop foot point and leading to its evaporation (Antonucci et al. 1982, 1984;
Fisher et al. 1985).

The thick-target model implies a causal connection between non-thermal
(microwave and hard X-ray) and thermal (soft X-ray and other) 
emissions, known as ``Neupert effect.''
This is the correlation between the time-integrated non-thermal emission 
and the instantaneous thermal emission (Neupert 1968; Dennis $\&$ Zarro 1993; Veronig et al. 2002): as the instantaneous non-thermal radiation is a proxy for the amount of evaporating plasma, its time-integral should correlate with the total amount of thermal plasma, and thus radiation being observed in soft X-rays, assuming that the time scale for cooling of the coronal plasma is much longer than the impulsive phase.

While the thick target paradigm has known difficulties by some authors, it 
has succeeded in explaining the emission observed in a number of very 
energetic stellar flares. In particular, the observed soft X-ray light 
curves of very large stellar events, lasting up to several days have been 
reproduced very well by numerical simulation of confined events in large 
coronal loops (Favata at al., 2005). The solar flaring mechanism seems then to be 
at work also in the most energetic stellar events, which can have peak 
energies of 4 to 5 orders of magnitude larger than solar events.

In the present paper we address two questions: first, how does the hard 
X-ray component in solar flares (dominated by non-thermal radiation) 
correlate with the soft X-ray flux (dominated by thermal radiation),
{Dennis $\&$ Zarro 1993; Veronig et al. 2002; Matsumoto et al. 2005.} 
Second, if a correlation exists for solar flares could we also 
extend it to the case of the most intensive stellar flares? And by extension, 
can non-thermal emission be observed in stellar flares, with present, past 
or planned hard X-ray detectors?

{In contrast to a  previous similar analysis by Battaglia et al. (2005),
here we include the strongest X-class flares.
A comparison between the their and our results will be given in Sect.~5.}

The present paper is organized as follows: in Sect.~2 we give a
brief description of the instrumentation used and the data analysis; 
in Sect.~3 we present the results obtained for the case of solar flares;
in Sect.~4 we extend these results to the case of stellar flares.
Finally {in  Sect.~5 we discuss our results and
in Sect.~6 we summarize the main conclusions of the paper.}

%__________________________________________________________________
\section{Observations and data analysis}

RHESSI (Reuven Ramaty High-Energy
Solar Spectroscopic Imager) is a NASA Small Explorer,
launched on February 5, 2002, which combines for the
first time high-resolution imaging in hard X-rays
and $\gamma$-rays with high-resolution spectroscopy,
in a way that a  detailed energy spectrum
can be obtained for the resolved solar disk (see Lin et al. (2002) 
and references therein for full details of RHESSI).

The spectroscopy is achieved with nine
cooled high-purity germanium crystals positioned behind the nine
grid pairs of the telescope. These convert incoming
X-rays and $\gamma$-rays to charge pulses.
The detectors have two electrically-independent
segments:  a front one to measure hard X-rays
up to 200~keV and a rear one for energies up to 17~MeV,
with spectral resolution of about 1~keV up to
100~keV and 3~keV up to 1~MeV.

The RHESSI design features movable aluminum shutters 
to attenuate the fluxes of soft  X-rays, preventing saturation 
and pulse pile-up that would occur at high count rates. 
The two shutter states used (A1, thin, and A3, {both thick and thin in}),
are characterized by their half-response energies
around 17 and 27 keV respectively.
All of the analyses reported here are based on the simple diagonal 
elements of the spectral response matrix. While non-diagonal elements 
of the response matrix become important at the lower energies 
(i.e.\ bbelow 10 keV), we have only used, from RHESSI, data 
at $E \ge 20$ keV, which justifies the approach chosen.

%__________________________________________________________________

\subsection{Data selection}

Solar flares are classified on the basis of the GOES
(Geostationary Operational Environmental Satellite)
observations (e.g., Garcia 1994), which monitor solar X-ray emission without spatial resolution (together with solar particle fluxes, etc.). The flare classification is based on the logarithm of the flux in the 1.6--12.4 keV band, so that each flare is identified by a letter (as per Table~\ref{table0}) plus a numeric code indicating the subclass.
\begin{table}[h!]
\caption{The correspondence between the GOES class of a solar flare and the flux at Earth in the 1.6--12.4 keV band.}             
\label{table0}     
\centering                          
\begin{tabular}{c c}        
\hline\hline                 
GOES Class & Flux 1.6--12.4~keV, ${\rm W} \cdot {\rm m}^{-2}$\\
\hline
A  & $1.0\times 10^{-8}$ \\     
B  & $1.0\times 10^{-7}$ \\
C  & $1.0\times 10^{-6}$ \\
M  & $1.0\times 10^{-5}$ \\
X  & $1.0\times 10^{-4}$ \\
\hline
\end{tabular}
\end{table}

Our aim was to study whether a general correlation exists between the peak 
soft and hard X-ray fluxes in solar flares, and to verify whether such 
correlation can also be extended to the case of intense stellar flares. 
We have therefore studied solar events spanning as broad a range of 
intensities as possible, from the weakest flares in the C1 class up to 
the strongest available X class flares. 

{We fixed a total number of around 50 events to study, and
we tried to sample each class (C, M and X) evenly, with 15 events
per class. The criteria used to select the flares in our sample were:}

\begin{itemize}
\item{The flare must have good GOES data}

\item{The peak hard X-ray emission must be well observed (no RHESSI night or South Atlantic Anomaly passage near peak time)}

\item{The flare must be well defined and not occur during the decay phase of a larger
one}

\end{itemize}

The list of flares which compose our sample is found in Table~\ref{table1}.
Note that the quiescent emission has been subtracted from the values of 
the peak photon fluxes for all three bands of interest. 
The value of the quiescent emission has been determined by selecting 
an interval in the pre-flare phase, possibly nigh time, for the RHESSI data
and interpolated between the pre-flare and post-flare
values for the GOES data. 

Many good candidates were available for class C~or M~is events, as thousands 
of flares have been observed in these two classes. 
{The class C ad M events listed in Table~\ref{table1}
are therefore a representative sample, albeit by necessity somewhat arbitrary. On the other hand for the class X events the choice is limited, as the total 
number of class X events observed by RHESSI through 2005 is only 62.
Once our criteria are applied, 25 class X events are left, from which we 
selected our sample of 15 events.} The clear correlation between the peak 
soft and hard X-ray emission found (e.g.\ Fig.~\ref{F6}) confirms a 
posteriori that our events are representative and selected from a 
homogeneous parent sample.

\begin{table}
\caption{The list of flares used for our analysis. The time is the peak time
in the 12--35~keV band as provided by the RHESSI Experimental Data Center
(HEDC; Saint-Hilaire et al. 2002). The asterisks denote flares for which we 
were able to determine the peak emission in both the 20--40 keV and in the 
60--80 keV bands. The classification under ``GOES orig.'' is the original 
GOES class before subtraction of the quiescent emission, while the one under 
GOES is the one with the quiescent emission subtracted, which we used in our
 work.}             % title of Table
\label{table1}      % is used to refer this table in the text
\centering                          % used for centering table
\begin{tabular}{c c c c c c}        % centered columns (4 columns)
\hline\hline                 % inserts double horizontal lines
&Flare & Date & Time & GOES & GOES orig. \\    % table heading
\hline                        % inserts single horizontal line
1& 3040507   & 05/04/03 & 00:35:34 & C1.3 & C2.1\\
2& 2022420   & 24/02/02 & 23:16:14 & C2.7 & C4.0\\
%3& 5062703   & 27/06/05 & 08:47:26 & C3.2 & C3.3\\
3& 3040905   & 09/04/03 & 03:48:58 & C3.2 & C3.7\\
4& 2040808   & 08/04/02 & 03:03:22 & C4.0 & C5.0\\
5& 2021213   & 12/02/02 & 21:33:06 & C5.0 & C6.0\\
6& 3011601   & 16/01/03 & 01:06:50 & C5.2 & C5.7\\
7& 2110510   & 05/11/02 & 16:09:06 & C6.3 & C7.1\\
8& 5082925   & 29/08/05 & 21:53:14 & C6.4 & C6.5\\
9& 2043001  & 30/04/02 & 00:33:18 & C7.1 & C8.0\\
10& 2071810  & 18/07/02 & 23:15:46 & C7.9 & C8.6\\
11& 2080601   & 06/08/02 & 01:43:46 & C8.1 & C9.0\\
12& 5073009   & 30/07/05 & 05:16:46 & C9.4 & C9.5\\
13& 2071106  & 11/07/02 & 14:18:10 & C9.9 & M1.0\\
14& 2022128  & 21/02/02 & 18:16:30 & C9.9 & M1.2\\
15*& 5051603  & 16/05/05 & 02:41:26 & M1.4 & M1.5 \\
16& 2042408  & 24/04/02 & 21:55:30 & M2.0 & M2.0 \\
17& 2093005  & 30/09/02 & 01:49:38 & M2.3 & M2.4 \\
18*& 2071814  & 18/07/02 & 03:34:54 & M2.4 & M2.5 \\
19*& 2092926  & 29/09/02 & 06:38:30 & M3.0 & M3.1 \\
20& 5091103  & 11/09/05 & 02:34:30 & M3.3 & M3.5 \\
21*& 2022001  & 20/02/02 & 09:58:02 & M4.7 & M4.9 \\
22*& 2072905  & 29/07/02 & 02:37:06 & M5.0 & M5.3 \\
23*& 2082267  & 22/08/02 & 01:52:38 & M5.6 & M5.9 \\
24*& 2040413  & 04/04/02 & 15:29:58 & M6.2 & M6.4 \\
25*& 2052010  & 20/05/02 & 10:52:50 & M6.2 & M6.5 \\
26*& 4071427  & 14/07/04 & 05:22:06 & M6.3 & M6.4 \\
27*& 4071301  & 13/07/04 & 00:15:38 & M6.7 & M6.8 \\
28*& 5082502  & 25/08/05 & 04:39:06 & M7.0 & M7.0 \\
29*& 2111802  & 18/11/02 & 02:07:30 & M7.9 & M8.0 \\
30*& 2041014  & 10/04/02 & 12:28:30 & M8.6 & M8.8 \\
%26& 4072242  & 22/07/04 & 00:30:02 & M8.8 & M9.1 \\
31*& 4081334  & 13/08/04 & 18:11:42 & X1.02 & X1.1 \\
32*& 2080327  & 03/08/02 & 19:06:54 & X1.18 & X1.2 \\
33& 4022604  & 26/02/04 & 02:01:02 & X1.18 & X1.2 \\
34*& 3052920  & 29/05/03 & 01:03:34 & X1.17 & X1.2 \\
35*& 4071606  & 16/07/04 & 02:05:30 & X1.32 & X1.4 \\
36*& 5011911  & 19/01/05 & 08:16:06 & X1.1  & X1.4 \\
37*& 2083048  & 30/08/02 & 13:29:50 & X1.57 &X1.6 \\
38*& 4071514  & 15/07/04 & 18:23:18 & X1.65 &X1.7 \\
39*& 5011710  & 17/01/05 & 09:47:06 & X1.4  &X1.7 \\
40*& 2070352  & 03/07/02 & 02:12:34 & X1.8  &X1.8 \\
41*& 4071503  & 15/07/04 & 01:39:46 & X1.76 &X1.8 \\
42*& 4111002  & 10/11/04 & 02:10:22 & X2.6  &X2.6 \\
43*& 3110301  & 03/11/03 & 01:38:12 & X2.7  &X2.7 \\ 
44*& 2072013  & 20/07/02 & 21:14:46 & X3.4  &X3.4 \\
45*& 5012005  & 20/01/05 & 06:49:54 & X7.1  &X7.1 \\
46*& 3102929  & 29/10/03 & 20:46:24 & X10   &X10  \\
\hline                                  
\end{tabular}
\end{table}

\subsection{Data analysis}

For each flare in the list we produced light curves in a number of energy 
bands from the RHESSI data, using the RHESSI software (Schwartz et al., 2002) 
with a 
time bin of one rotation period of the RHESSI instrument collimator
($\simeq$ 4 seconds). From the light curves in the relevant energy bands 
we determined the photon fluxes at the peak of the flare in the 20--40 keV 
and 60--80 keV bands.
We performed a similar operation for the peak energy flux in the 1.6--12.4 
keV band from GOES data.

While the RHESSI passband in principle extends {down to 3 keV}, we prefer 
to use the GOES data for the soft (mostly thermal) emission as the analysis of 
the RHESSI data in these bands is complicated by the presence of the moving 
attenuators for the more energetic events, as well as by the more complicated 
response of the detectors.

We have chosen to compare the fluxes at the peak of the events, rather than e.g.\ total energy released in each band, as this is a quantity which can be determined more reliably in stellar flares, also for events with a modest signal-to-noise, for which one cannot reliably follow the flare decay (during which significant amounts of energy are still released).

To make our analysis as non-parametric as possible, we have decided not to rely on spectral fits to the RHESSI data. To determine photon fluxes from the raw counts measured by the RHESSI detectors we have used the ``semi-calibrated'' data produced by the RHESSI data pipeline. As described
above this implies using of the diagonal elements from the response matrices of the detector. For this analysis to be reliable, the following conditions have to be satisfied:

\begin{enumerate}

\item The flare counting rate over the background should be at least ten times
the background rate for the energies considered.

\item Only energies substantially below 100~keV can be dealt with properly in the semi-calibrated analysis.
This is due of the increasing importance of 
Compton scattering at higher energies. 
This has driven the choice of adopting 80~keV as the upper limit on our 
high-energy band.

\item The fluxes below 10 keV are reliable only if 
no attenuators were in place. In the presence of attenuators the 3--10 keV
band can be dominated by germanium K-shell escape events.
This limitation in the RHESSI data has driven our choice to use data from GOES for the 1.5--12.4 keV band.

\end{enumerate}

We used only data from the front segments of the RHESSI detectors, 
as we focused on energies below 100 keV. We also discarded data 
from detectors~2 and~7 because of their bad resolution.

The RHESSI pipeline software automatically corrects for
front-segment ``decimation'' (a RHESSI telemetry-saving
feature) which produced loss of counts below the
decimation upper threshold energies. It also corrects for the presence of the attenuators.
The pile-up correction is disabled by default but we enable this for all
strong flares. 

\begin{figure}
\begin{center}
\includegraphics[width=9.00cm,height=8.0cm,clip=true]{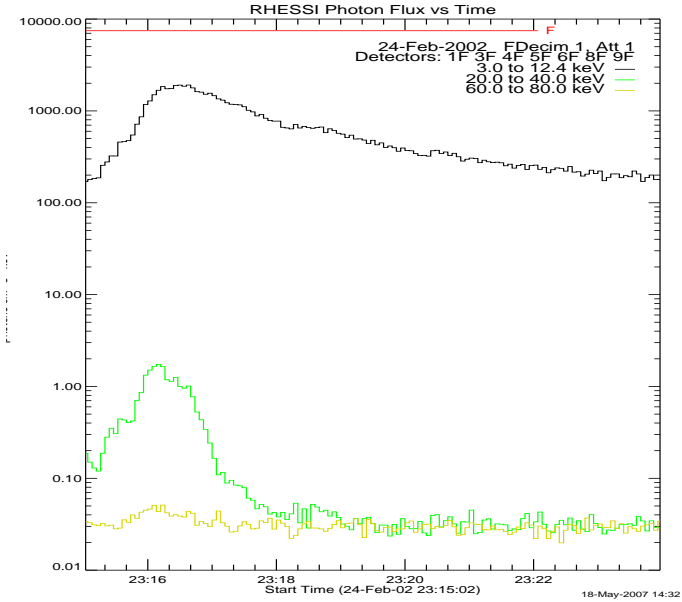}
\caption{The light curves for flare~2 from Table~\ref{table1}, which has GOES class~C2.7. Fluxes are in units of ph/$({\rm cm^{2}}\,{\rm s}\,\,{\rm keV}$).}
\label{F1}
\end{center}
\end{figure}

\begin{figure}
\begin{center}
\includegraphics[width=9.00cm,height=8.0cm,clip=true]{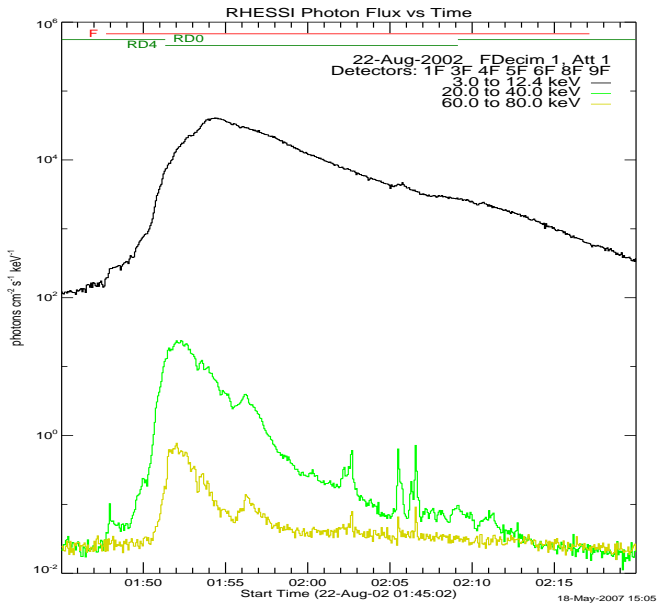}
\caption{The light curves for flare~23 from Table~\ref{table1}, which has GOES class~M5.6. Fluxes are in units of ph/$({\rm cm^{2}}\,{\rm s}\,\,{\rm keV}$).}
\label{F2}
\end{center}
\end{figure}

Figs.~\ref{F1} and~\ref{F2} show the light curves of 2 flares in class C and M
respectively. In addition to the light curves in the 20--40 and 60--80 keV bands {we also plot the emission in 3.--12.4 keV band from RHESSI, which for 
the weaker flares is not affected by the presence of attenuators and thus it
is close to the GOES band 1.6-12.4}. We have however not used this in our 
analysis (relying on the GOES data). Note how the emission in the 60--80 keV band 
is barely visible (if at all) for the C class flare, while it becomes clearly
 present in the M class event. In both events the peak emission in the 
20--40~keV band 
precedes the one in the soft component  as expected if the hard emission 
is associated with the impulsive, non-thermal phase. {The M-class event,
by comparison of the 60-80 and 20-40 keV RHESSI bands, shows the so-called 
``soft-hard-soft'' spectral evolution as expected} ({Parks $\&$ Winckler, 1969}; 
Hudson $\&$ F\'arn\'ik, 2002; Fletcher $\&$ Hudson, 2002; {Grigis $\&$ Benz, 2005}). 

\section{Scaling laws for solar flares}

\begin{center}
\begin{figure}[h!t!]
\includegraphics[width=9.0cm,height=7.0cm,clip=true]{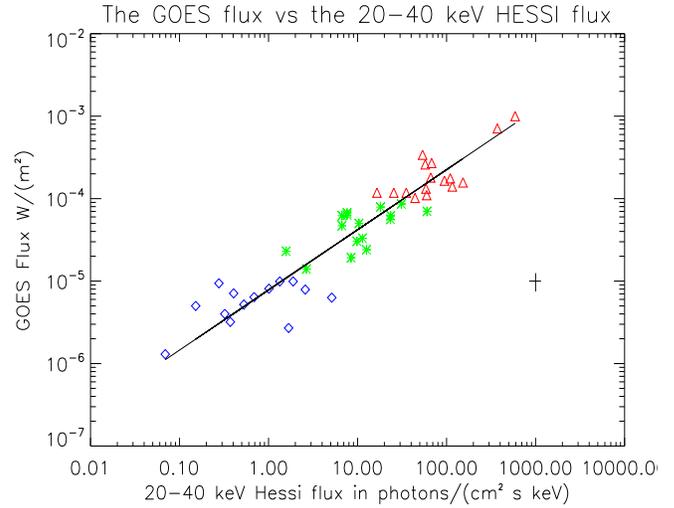}
\caption{The peak emission in the 1.6-12.4~keV band from GOES data as a function of the peak emission in the 20-40~keV band, from RHESSI data, for all flaring events in our sample. Blue diamonds, green stars and red triangles correspond to C, M and X 
class respectively. A representative error bar is shown in the lower right 
part of the plot.}
\label{F6}
\end{figure}
\end{center}

Fig.~\ref{F6} shows the correlation between the peak emission in the 
1.6-12.4~keV band, as observed by GOES, and the RHESSI 20-40~keV band 
peak emission, for all flares in our sample (Table~\ref{table1}). 
{The representative error 
bar on the RHESSI data has been estimated from the data itself
combining the error on the
peak flux and on the background. For the peak we considered the 
fluctuations around the peak in a short binning of 4 seconds.
Typical values were 15$\%$ for the 20-40 keV band and 10$\%$ for
the 60-80 keV band. For the background we took the peak to
peak excursion and for the GOES band we estimated an error 
of 20$\%$.}

The two quantities are well correlated, and a scaling law in the form of a power law can be derived, applying the bisector regression method as described
in Isobe et al. (1990):
\begin{equation}
F_{\rm G}\sim  7.83 \times 10^{-6}~F_{(20-40)}^{0.73}
\label{EQ1}
\end{equation}
where $F_{\rm G}$ is the GOES peak flux 1.6-12.4~keV band in units of ${\rm Watt}/{\rm m}^2$ and $F_{20-40}$ is the peak flux in the 20-40 keV band, in ${\rm ph}/({\rm cm}^{2}\, {\rm s}\,\, {\rm keV})$. 
The slope for the power law is $0.73 \pm 0.04$
and the intercept is $(7.83 \pm 0.98) \times 10^{-6}$. This correlation 
holds over more than 3 orders of magnitude in GOES peak flux.
{An analysis on the comparison between this scaling law and the 
one found in Battaglia et al. (2005) is included in the Section 5.}

\begin{center}
\begin{figure}[h!]
\includegraphics[width=9.0cm,height=7.0cm,clip=true]{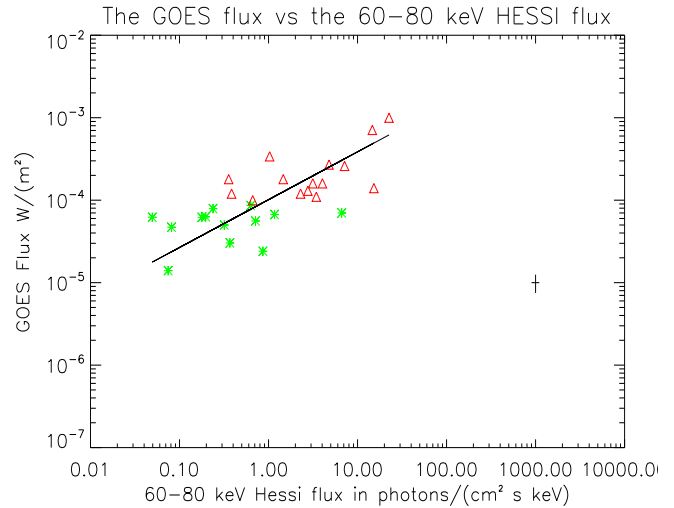}
\caption{The peak emission in the 1.6-12.4~keV band from GOES data as a function of the peak emission in the 60-80~keV band, from RHESSI data, for { all flare events with detectable 60-80 keV fluxes}. Green stars and red triangles correspond to M and X 
class events respectively. A representative error bar is shown in the lower right part of the plot.}
\label{F7}
\end{figure}
\end{center}

In Fig.~\ref{F7} we present the same type of analysis as in Fig.~\ref{F6}
but considering peak emission in the 60-80~keV band.
Note that the number of events here is less than for the 20-40 keV band, 
as for the weaker flares no emission is detected in the 60-80 keV band. 
The flares for which a 60-80 keV peak flux has been determined are indicated 
with an asterisk in Table\,\ref{table1}.

Although the spread is more significant than in the case of the 20-40 keV flux, a correlation between soft and hard peak fluxes is still well visible. We derived, using the same approach as for Eq.~\ref{EQ1}, a power law
\begin{equation}
F_{\rm G}\sim 1.02\times 10^{-4}\,\,F_{(60-80)}^{0.58}
\label{EQ3}
\end{equation}
where the units are the same in Eq.~\ref{EQ1} and the slope is $0.58 \pm 0.07$ and the intercept $(1.01 \pm 0.13) \times 10^{-4}$. 

%__________________________________________________________________

\subsection{Thermal contribution to the hard band flux}

We have considered the total peak flux in each band, regardless of its origin. To determine which fraction of the observed flux is of thermal origin, we have developed a procedure to subtract from the total peak flux observed in the 20-40 and in the 60-80 bands the thermal component, as a function of the peak flare temperature, using a set of model spectra. This is necessary to compare the predictions of our scaling laws with the observations of intense stellar flares, where the observed plasma temperatures are much higher than for solar flares and thermal emission can  be significant also in these bands. 

\begin{center}
\begin{figure}[h!]
\includegraphics[width=9.0cm,height=6.5cm,clip=true]{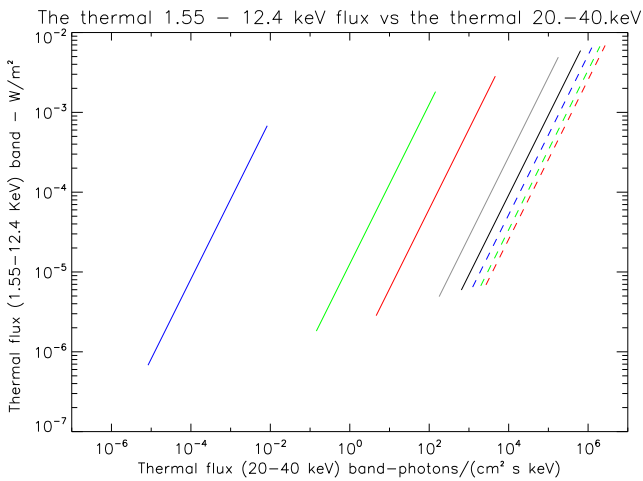}
\caption{The flux in the 1.6-12.4 keV band as a function
of the flux in the 20-40 keV band for a MEKAL thermal models, for temperatures of
1, 2, 3, 6, 9, 12, 16, and 20~keV from left to right, which correspond
respectively to $\sim$12, 23, 35, 70, 104, 139, 186 and 232~$\times$~10$^{6}$K.}
\label{F8}
\end{figure}
\end{center}

To do this, we have simulated a set of thermal spectra and determined the 
relationship between the flux in the GOES 1.6-12.4 keV band and the flux 
in the 20-40 and 60-80 keV bands as a function of the peak temperature of 
the flare. We used the XSPEC package (Schwartz 1996; Smith et al. 2002) 
with a MEKAL thermal model, a plasma emission code which implements
the optically-thin collisional ionization equilibrium emissivity 
model described in Mewe et al. (1995).
For different temperatures we determined the value of the fluxes in the 
different bands of interest, 1.6-12.4, 20-40 keV and 60-80 keV.
In Fig.~\ref{F8} we plot the flux (as given
by the model) in the 1.6-12.4 keV versus the flux in the 20-40~keV band for
different temperatures ranging from 1~keV to 20~keV. 

The relationship between the flux in the two bands is obviously linear, 
with a proportionality constant which depends on the temperature. 
Writing the relation as
\begin{equation}
F_{(20-40)} = m\cdot F_{\rm G}
\label{EQ3bis}
\end{equation}
and using the same units as before, i.e.\ $F_{\rm G}$ in ${\rm W}/{\rm m}^2$ 
and $F_{(20-40)}$ in ${\rm ph}/({\rm cm}^2\,{\rm s}\,\, {\rm keV})$, 
the values of $m$ for different temperatures are given in Table~\ref{table2}, 
for both the 20-40 and the 60-80 keV band.
\begin{table}[h!]
\caption{The values of $m$ in giving the proportionality between the thermal flux in the 20-40 and 60-80 keV bands ($F_{(20-40)}$ and $F_{(60-80)}$ and the thermal flux in the GOES 1.6-12.4 keV band ($F_{\rm G}$), with $F_{\rm G}$ in ${\rm W}/{\rm m}^2$ and $F_{(20-40)}$ and $F_{(60-80)}$ in ${\rm ph}/({\rm cm}^2\,{\rm s}\,\, {\rm keV})$.}   
\label{table2}      % is used to refer this table in the text
\centering                          % used for centering table
\begin{tabular}{ccc}        % centered columns (4 columns)
\hline\hline                 % inserts double horizontal lines
Temp. (keV) & $m$, 20-40 keV & $m$, 60-80 keV\\
\hline
2  & $7.93\times 10^{4}$ & $4.10\times 10^{-5}$\\
3  & $1.62\times 10^{6}$ & 0.69\\
6  & $3.62\times 10^{7}$ & $1.27\times 10^{4}$\\
9  & $1.08\times 10^{8}$ & $3.63\times 10^{5}$\\
12  & $1.92\times 10^{8}$ &$1.98\times 10^{6}$\\ 
\hline
\end{tabular}
\end{table}

While we cannot attribute a peak temperature individually to each flare 
in our sample, Feldman et al. (1995) found a tight correlation between peak 
flare intensity (up to GOES class X1) and peak temperature, allowing us to 
attribute an average temperature for some reference classes of flares.
For each of these temperatures we calculate the thermal contribution
with XSPEC and the corresponding total predicted flux obtained reversing
the scaling law in Eq.\ref{EQ1} which gives the following expression:
\begin{equation}
F_{(20-40)}\sim 9.908 \times 10^{6} F_{\rm G}^{1.37}
\label{EQ5}
\end{equation}
in the same units used above.

We computed the  ratio between the peak total flux (obtained from 
Eq.~\ref{EQ5}) and the average peak thermal flux in the 20-40 keV band. 
The results are shown in Table~\ref{table2bis}.
\begin{table}[h!]
\caption{The ratio between the peak thermal and non-thermal flux in the 
20-40~keV band as a function of the flare GOES class. The first column 
reports the class of flares, 
the second column the average peak temperature as derived from Feldman 
et al. (1995), the third column the peak thermal flux 
(in units ${\rm ph}/{\rm cm}^2\,{\rm s}\,\,{\rm keV}$) from the MEKAL 
models, and the last column the ratio between the total peak flux derived 
by the  scaling law in Eq.~\ref{EQ5} and the average peak thermal flux.}             
\label{table2bis}     
\centering                          
\begin{tabular}{cllr}        
\hline\hline                
Class & Average $T$ (keV) & $F_{\rm th (20-40)}$ & $F_{(20-40)}/F_{\rm th (20-40)}$\\
\hline
C1 & 1.30  & 0.00058  &$\sim$ 100 \\
C4 & 1.47  & 0.02  &$\sim$ 24 \\
M1 & 1.64 & 0.10  &$\sim$ 14\\
M4 & 1.70 & 0.54 &$\sim$ 17\\
X1 & 2.00 & 7.92  &$\sim$ 4\\
\hline
\end{tabular}
\end{table}

{We can see that the thermal contribution to the peak flux in 
the 20-40 keV band is negligible in the less intense flares,
and can be relevant in the 20-40~keV band only for the more 
intense events and and for temperatures close to the maximum 
that is observed on the Sun}.

Using the data in Table~\ref{table2} and Eq.~\ref{EQ1} for
a temperature of around 2~keV, for instance, we find that a value 
above $F_G \sim 3\times 10^{-6}\,\,{\rm W}/{\rm m}^2$  is required 
in order to have a total predicted $F_{(20-40)}$ emission higher 
or equal to the thermal component.
This means that temperatures of the order of 20~$\times$~10$^{6}$K 
could be reached most likely for flares in M and X class, which 
is consistent with the result found by Feldman et al. (1995).

%{Taking an M5 flare and assuming always a temperature
%around 2 keV, the flux in the 20-40~keV band (always deduced from 
%Eq.~\ref{EQ1}) will be around three times the thermal one; 
%going to a flare of class X5, for the same temperature, 
%the total emission will be eight times the thermal one. 
%This means that keeping constant the temperature, the 
%non-thermal component increases rapidly with the 
%intensity of the flare.}

\begin{center}
\begin{figure}[h!]
\includegraphics[width=9.0cm,height=7.0cm,clip=true]{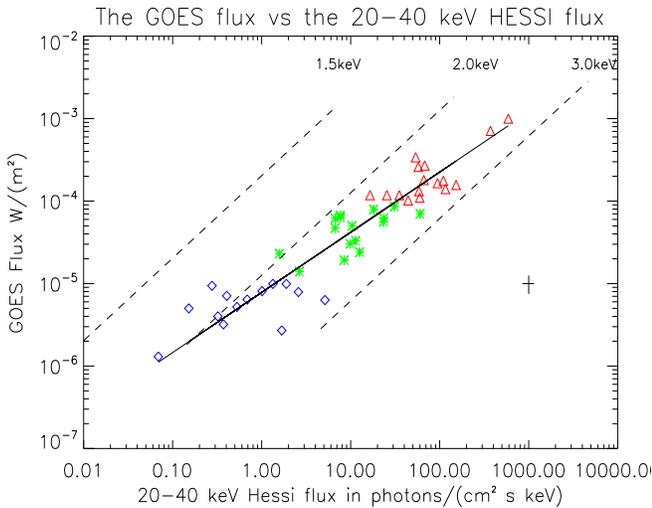}
\caption{The same data as in Fig.\ref{F6} with three additional
dashed lines showing the thermal flux in the GOES band 
as a function of the thermal flux in the 20-40~keV band
for peak flare temperatures of 1.5, 2.0 and 3.0 keV (spanning the range 
observed for solar flares from the C1 to the X10 class).}
\label{F9}
\end{figure}
\end{center}

In Fig.~\ref{F9} we show the same data as in Fig.~\ref{F6} for the peak 
emission during our sample of solar flares, with superimposed the relation 
between the thermal flux in the GOES band versus the thermal flux in the 
20-40~keV band for a peak temperature of 1.5, 2.0 and 3.0 keV, spanning 
the range of temperatures observed in solar flares from the C1 to the X10 
classes. {The class X flares are all located between the loci for 
the thermal fluxes for plasma at 2.0 and 3.0 keV (peak temperatures 
which are not exceptional for class X events). If the 
temperatures associated to our observed flares were lower than 2 keV,
the fluxes would be totally dominated by non-thermal effects.
For temperatures around 2.5 keV a part of the emission in this band, 
for the intense solar flares, should be given by the tail of the 
thermal spectrum.}

\begin{center}
\begin{figure}[h!]
\includegraphics[width=9.0cm,height=7.0cm,clip=true]{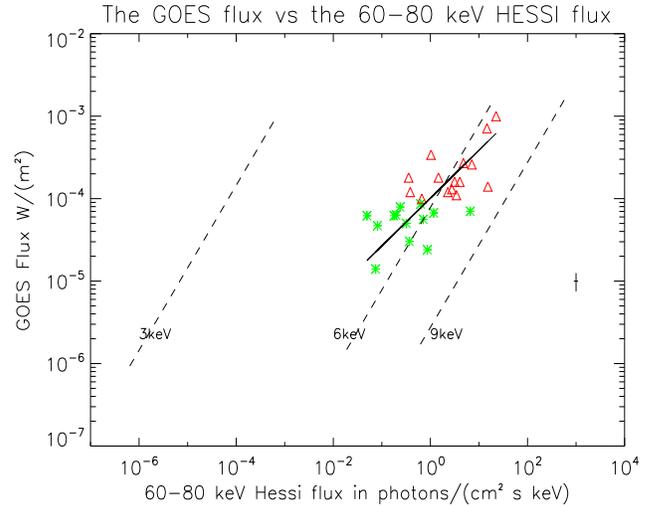}
\caption{The same as Fig.\ref{F7} with three additional
dashed lines showing the thermal flux in the GOES band 
as a function of the thermal flux in the 60-80 keV band
for plasma at 3.0, 6.0, 9.0 keV.}
\label{F10}
\end{figure}
\end{center}

In Fig.~\ref{F10} we show the same data as in Fig.~\ref{F9} for the 
60-80~keV band. In this band the observed emission is entirely non-thermal: 
to contribute to the observed flux the thermal plasma would need to have 
temperatures $\ga 6$~keV, that is much higher than what is observed in solar events. 

\section{Stellar flares}

{In the previous section we have established a scaling law between 
the peak flare emission in the GOES, 20-40 keV and 40-60 keV bands in solar 
flares, including the strongest X-class. In this section we  go
further and determine whether the same law extends to the emission from 
intense stellar flares. This is a step beyond all previous studies 
which provides interesting new results.}
 
For this purpose we have compared the extrapolation of the solar scaling 
law to the fluxes observed in stellar events with the few available 
broad-band data from intense stellar flares. The only observatory which 
has to date performed broad-band observations of stellar flares, from 0.1 
up to $\simeq 100$ keV has been {\it Beppo}SAX  (Boella et al. 1997a). We 
have analyzed the data from the MECS  (1.6-10~keV) (Boella et al. 1997b) 
and PDS (15-300~keV)  (Frontera et al. 1997) instruments for the flares 
observed on UX Ari, (Franciosini et al. 2001),  Algol, (Favata et al. 2001) 
and AB Dor (2 events, Maggio et al. 2000). For the purpose of the present 
paper we needed to derive the peak fluxes in the 3 bands of interest, 
distinguishing the total, thermal and non-thermal fluxes. 

\begin{center}
\begin{figure}[h!]
\includegraphics[width=9.0cm,height=7.0cm,clip=true]{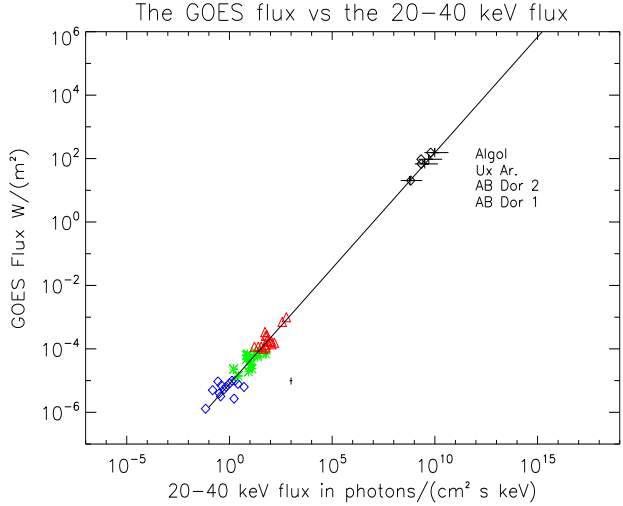}
\caption{The scaling law for the peak emission during solar flares,
 extrapolated to the regime of intense stellar flares, showing the relation 
between the peak flux in the GOES band versus the peak flux in the 20-40~keV 
band. The solar correlation (Eq.~\ref{EQ5}) is shown, extrapolated to the 
peak GOES flux of stellar flares. 
The diamonds, stars and triangles represent the data from C, M and 
X~class solar flares on the bottom left side. The black crosses show the 
corresponding location for the flares observed 
in the named stars in the top right side, using their peak flux in the 
GOES band, with the horizontal error bar representing the uncertainty 
induced by the uncertainty in the slope of the best fit correlation.
The diamonds on the left side of the data point for each star 
show the relation between the non-thermal peak flux as derived from the 
{\it Beppo}SAX data.}
\label{F13}
\end{figure}
\end{center}

\begin{center}
\begin{figure}[h!]
\includegraphics[width=9.0cm,height=7.0cm,clip=true]{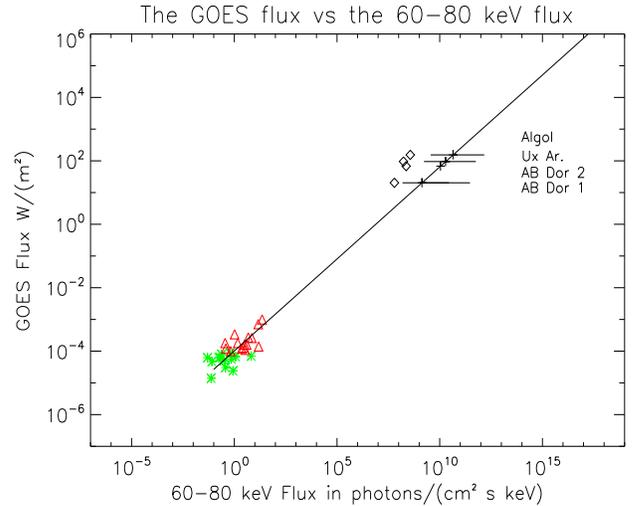}
\caption{The same as Fig.~\ref{F13} but for the 60-80~keV band.}
\label{F14}
\end{figure}
\end{center}

We extrapolated the solar scaling laws, and used the peak flux in the
GOES band (determined from our analysis of the {\it Beppo}SAX data) for 
each stellar flare to determine the predicted flux in the 20-40 and 60-80 
keV band, using Eq.~\ref{EQ5} and the following one:
\begin{equation}
F_{(60-80)} \sim 7.680 \times 10^{6}\, F_{G}^{1.72}.
\label{EQ6}
\end{equation}
The results are shown 
in Fig.~\ref{F13} for the 20-40 keV band and in Fig.~\ref{F14} for the 60-80
 keV band (where 
the stellar fluxes have been rescaled to a 1 AU distance, for ease of 
comparison). The horizontal error bars for the stellar events represent the 
uncertainties in the 20-40 keV flux resulting from the uncertainty in the 
slope of the best-fit correlation for the solar flares.

Table~\ref{table3} reports the results of this analysis for the 20-40 keV 
and 60-80 keV bands, respectively, giving, for each flare, the best-fit 
peak flux as deduced by the scaling law (Eq.~\ref{EQ5} and Eq.~\ref{EQ6}, 
scaled to the respective distance of each star), with the uncertainty 
induced by the uncertainty in the best-fit scaling law. The best-fit for 
the total peak flux as measured on the {\it Beppo}SAX spectra is also reported.

{We have performed a new analysis of the data, 
and we tried different possible models to fit the spectra. 
In particular we investigated if the presence of a non-thermal power law 
component could have produced a better fit, especially for the PDS
instrument. The results obtained showed that the precision of the data at
hand is not sufficient to draw any conclusion. A multi-temperature fit or
a one-temperature plus a power-law fit are equally statistically acceptable.
This means that we cannot exclude for sure
the fact that this component has already been observed with {\it Beppo}SAX
but not even demonstrate its presence. A recent study by Osten et. al. 2007 
also investigates possible non-thermal emission in a stellar flare.} 

\begin{table}[h!]
\begin{minipage}[t]{\columnwidth}
\caption{For each of the 4 intense stellar flares observed by {\it Beppo}SAX, the peak flux  in the 20-40~keV band and 60-80~keV as predicted by the extrapolation of the solar scaling law (column 2, Eq.~\ref{EQ5} and Eq.~\ref{EQ6}) is given, together with its uncertainty (column 3). In column 4 the peak flux determined from the observations (using a fit including a power-law component) is given. Units are ${\rm ph}/{\rm cm}^2\,{\rm s}\,\, {\rm keV}$.}            
\label{table3}      
\centering  
\renewcommand{\footnoterule}{}                 
\begin{tabular}{cccc}        
\hline\hline                 
Event & $F_{(20-40)}$ & Range & $F_{(20-40)}$ \\
      &  scaling law & & observed \\
\hline
Ux Ari\footnote{ Franciosini et al. (2001)} &$0.47\times 10^{-4}$ &  $[0.15,2.12]\times 10^{-4}$ & $0.20\times 10^{-4}$ \\  
Algol\footnote{Favata et al. (1999)} & $2.86 \times 10^{-4}$ &$[0.89,13.3]\times 10^{-4}$&$1.86\times 10^{-4}$ \\
AB Dor, 1\footnote{Maggio et al. (2000)}&$0.64 \times 10^{-4}$ &$[0.23,2.5]\times 10^{-4}$& $0.92\times 10^{-4}$\\
AB Dor, 2\footnote{Maggio et al. (2000)}& $3.35\times 10^{-4}$ &$[0.11,14.6]\times 10^{-4}$& $2.28\times 10^{-4}$ \\
\hline
    & $F_{(60-80)}$ & Range & $F_{(60-80)}$ \\
          &  scaling law & & observed \\
\hline
UX Ari& $ 1.81 \times 10^{-4}$ & $[1.60\times 10^{-5},0.0053]$ & $0.16 \times 10^{-5}$\\
Algol & $ 1.29 \times 10^{-3}$ & $[1.06\times 10^{-4},0.043]$ &$ 1.34\times 10^{-5}$\\
AB Dor, 1  & $ 1.42 \times 10^{-4}$  &$[1.61\times 10^{-5},0.0039]$ &$0.23 \times 10^{-5}$\\
AB Dor, 2  & $ 1.14 \times 10^{-3}$ &$[1.07\times 10^{-4},0.031]$& $ 2.42 \times 10^{-5}$\\

\end{tabular}
\end{minipage}
\end{table}

From the upper part of Table~\ref{table3} it is evident that the extension of the correlation observed for solar flares between the peak emission in the GOES band and the emission in the 20-40 keV band fits rather well the data for the intense stellar flares observed with {\it Beppo}SAX. In all cases, the observed peak flux in the 20-40 keV band is within the range of values predicted from the extension of the solar relationship. Given that this entails an extrapolation over the solar GOES band peak flux by four orders of magnitude, it is remarkable that the simple solar scaling law succeeds in predicting the 20-40 keV band flux also for stellar flares, also given that their peak plasma temperatures are much higher.

The lower part of Table~\ref{table3} shows on the other hand that the 
extension of the solar correlation for the peak flux in the 60-80 keV 
band does not do a good job at predicting the peak flux in stellar events. 
In all cases, the predicted 60-80 keV peak flux is much higher than the 
observed value, by one to two orders of magnitude.

The correlation derived for the peak 60-80 keV flux in the solar case has 
a much steeper slope than the one derived for the peak 20-40 keV flux. 
In particular, when extrapolating toward very intense flares, the use of
 the solar correlations would predict that around class X100\,000 
(the most intense stellar flares are of class $\simeq $X1\,000\,000) 
the flux in the 60-80 keV band would exceed the one in the 20-40 keV band, 
leading to an ``inverted spectrum'' type of situation, contrarily to all 
available observational constraints. 

{The results in Table~\ref{table3} are obtained considering a model
with an additional power law, whereas in Table.~\ref{table6} we give
the values of the peak fluxes considering just the contribution
from the thermal component.}

{From this table we can immediately deduce that in
the 20-40 keV band almost all emission is of thermal origin while 
in the 60-80 keV, as for the Sun, it is the non thermal contribution 
which dominates.}
\begin{table}[h!]
\begin{minipage}[t]{\columnwidth}
\caption{For each of the 4 intense stellar flares observed by {\it Beppo}SAX, 
the thermal contribution to the peak flux  in the 20-40~keV band and 60-80~keV.Units are ${\rm ph}/{\rm cm}^2\,{\rm s}\,\, {\rm keV}$.}            
\label{table6}      
\centering  
\renewcommand{\footnoterule}{}                 
\begin{tabular}{ccc}        
\hline\hline                 
Event & $F_{(20-40)}^{th}$ &  $F_{(60-80)}^{th}$ \\
\hline\hline    
Ux Ari &$0.24\times 10^{-4}$ &  $0.64 \times 10^{-7}$ \\  
Algol & $1.2 \times 10^{-4}$ &  $ 9.6 \times 10^{-7}$\\
AB Dor, 1&$0.35 \times 10^{-4}$ &$0.43\times 10^{-7}$\\
AB Dor, 2& $1.4\times 10^{-4}$ & $4.60 \times 10^{-7}$ \\
\hline
\end{tabular}
\end{minipage}
\end{table}

{As the extrapolation to the intense stellar flares
in this band of energy seems to not be in agreement with 
the data at hand, probably different mechanisms play 
a dominant role that becomes more evident for the very
intense flares at the highest energies and that can
justify the discrepancy in the extrapolation of the
solar scaling law to stellar flares in the 60-80 keV 
band.}

Beyond the ``standard paradigm'' for solar flares, which we 
have identified with the thick-target model of Brown (1971),
other possibilities exist. 
In particular coronal hard X-ray sources of the type observed
by Frost and Dennis (1971) and Hudson (1978) now seem
to be almost routinely observed in major events by RHESSI
(Krucker et al., 2007).

\section{Discussion}

{The correlation law in solar flares 
represented by our Eq.~\ref{EQ1} can be compared to the
result found in Battaglia et al. 2005. They studied a sample of RHESSI 
flares, by performing 
spectral fits to separate the thermal emission from the non-thermal one 
(modeled as a power law), and restricting themselves to flares up to GOES 
class M5.9 (no X-class events were considered). They correlate the  
flux in the GOES band with the non-thermal monochromatic flux at 35 keV, 
find the scaling law
\begin{equation}
F_{\rm G} = 1.8 \times 10^{-5} F_{35}^{0.83}
\label{EQ2}
\end{equation}
where $F_{\rm G}$ is the flux in the GOES band and $F_{35}$ is the
non-thermal flux at 35~keV as obtained through individual spectral 
fits, in the same units as in Eq.~\ref{EQ1}.

Considering the different approach adopted here, the two scaling laws 
are quite consistent, especially regarding the slope, in agreement with the 
expectation that the emission in the GOES band will be dominated by 
thermal emission for the complete range of temperatures spanned by 
solar flares, while at 35 keV the emission observed will be mostly of 
non-thermal origin.

{On the other hand caution should be used in comparing
directly Eq.~\ref{EQ1} and Eq.~\ref{EQ2} where different
physical quantities appear. We used integrated fluxes 
in the 20-40 keV and 60-80 keV bands over all the energy,
whereas Battaglia et al. used a monochromatic spectrum
at 35 keV. This may justify a steeper slope in Eq.~\ref{EQ2} than
the one derived by us in Eq.~\ref{EQ1}.}

We would like to stress the fact that we do not make spectral fits,
which is also a major difference with the work by Battaglia. 
We decided to proceed in a non-parametric way because of the
difficulty to estimate the low-energy cut-off of the power
law from the data. This can produce a significant uncertainty in the
determination of the thermal component. {Data itself cannot 
constraint well the two components and then the derived parameters 
could be very uncertain.}

In Fig.~\ref{F16} we represent the same as in Fig.~\ref{F14} with 
an additional fit obtained combining the solar data and the
stellar observations. 
\begin{center}
\begin{figure}[h!]
\includegraphics[width=9.0cm,height=7.0cm,clip=true]{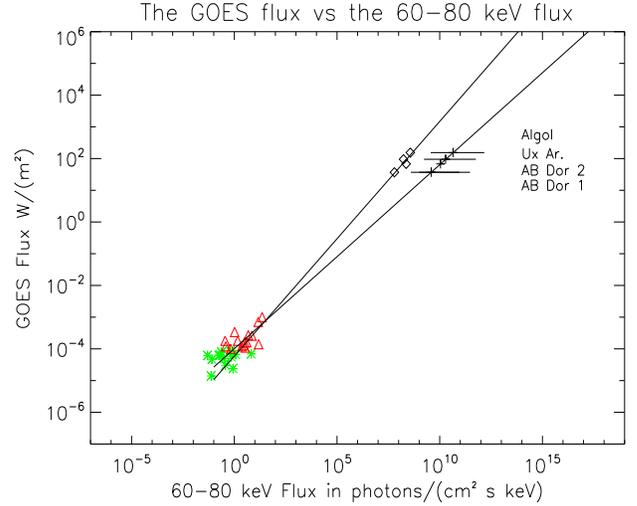}
\caption{The same as in Fig.~\ref{F14} with an additional fit 
obtained combining the solar RHESSI data and stellar flares
observations.}
\label{F16}
\end{figure}
\end{center}
This produces a scaling law 
\begin{equation}
F_{(60-80)}\sim 6.20 \times 10^{5} F_{\rm G}^{1.37}
\label{EQ7}
\end{equation}
which coincidentally has the same slope as in Eq.~\ref{EQ5}
for the 20-40 band. 
We remind that the scaling law for the 60-80 keV band 
has been deduced from a limited number of flares in
Table~\ref{table1} as the less intense events have a 
negligible emission at these high energies.

If future additional points would confirm the relation 
in Eq.~\ref{EQ7}, this would imply an universal slope across
more than six orders of magnitude in flare strength,
which will be a quite remarkable result.}

%__________________________________________________________________

\section{Conclusions}

We have studied the correlation between the peak flux emitted in three 
different energy bands (1.6-12.4 keV, or ``GOES'' band, 20-40 keV and 
60-80 keV) for a number of solar flares spanning a broad range of peak 
flux in the GOES band (or ``GOES class''). In the solar case, the GOES-band 
flux is almost completely dominated by the thermal emission of the flaring 
plasma, while the 60-80 keV flux is entirely due to non-thermal emission. 
This is likely due to the accelerated electrons which in the thick-target 
model of flares cause the evaporation of the plasma which is responsible 
for the thermal emission. The 20-40 keV emission contains contributions 
from both the tail of the thermal spectrum and the non-thermal, power 
law spectrum, with the former increasing with the GOES class, which in solar 
flares is well correlated with the peak temperature of the flare.

The analyzed sample of solar flares spans 3 orders of magnitude in GOES 
class (from C1 to X10), and both the 20-40 keV and the 60-80 keV peak 
flux are well correlated with the GOES band peak flux. The good correlation 
for the 60-80 keV band (which in the solar case is purely non-thermal in 
origin) may be construed to imply support for a ``thick-target'' type of 
framework, in which the thermal emission is linked to the non-thermal one 
by a causal relationship. 

We have then studied the applicability of these scaling laws to the case of intense stellar flares. Our sample is small, limited by the availability of the data: {\it Beppo}SAX has been the only X-ray observatory with the broad-band response needed for these observations, and only 4 events have been detected in the hard bands. In the future, additional stellar flares may be observed with Suzaku, and, later with the proposed Simbol-X hard X-ray telescope.

The extrapolation of the scaling laws derived for solar flares to the case 
of intense stellar flares (an extrapolation of 5 orders of magnitude in GOES
 peak flux) results in a quite remarkable good agreement between the 
predicted and observed 20-40 keV flux, while it over-predicts by some 
two orders of magnitude the 60-80 keV flux.

{Future observations, increasing the statistic of intense
flares, should allow to clarify the origin of this 
discrepancy. If a power law like the one in Eq.~\ref{EQ7} is 
found, the extrapolation would be possible for both bands and
the two scaling laws would be defined by the same slope, which
would be an even more remarkable result.}

%__________________________________________________________________

\begin{acknowledgements}
The contribution of HSH and G.M to this work was supported by project ISHERPA,
funded by the European Commission as contract No MTKD-CT-2004-002769.
The contribution of HSH was also supported by NASA under NAS 5-98033. 
C.I wishes to thank Tim Oosterbroek and Solen Balman for the useful discussions.

\end{acknowledgements}

\end{document}